**Title:** A strategic architecture for growing a space economy utilizing foundational space weather
**Short title:** The Anemomilos Plan
**Author:** W. Kent Tobiska, Space Environment Technologies, ktobiska@spacewx.com

## Abstract

We face unprecedented resource stresses in the 21st Century such as global climate disruptions, freshwater scarcity, expanding energy demands, and the threat of global pandemics. Historically, societies have relieved resource stress by increasing trade, innovating technologically, expanding territorially, regulating, redistributing, making alliances, creating new economic models, training new skills, as well as conducting war. Do we continue depleting our already strained resources leading to more regulation, redistribution, alliances, new economics, and war or do we grow our resources using innovation, expansion, new economics, and new skills? We present the argument for evolving space development using asteroid mining as the primary activity for frontier expansion aided by Low Earth Orbit (LEO), Moon, and Mars waystations. Forecast space weather is a necessary technology baseline for developing this pathway. All activity off Earth will require a fundamental knowledge of how the energetics of space will affect technological progress. We discuss the critical elements this space economy expansion, including technical feasibility and infrastructure development, economic and geopolitical viability complete with the US National Space Weather Program dialogue, ethical and legal considerations, and risk management. This discussion helps us understand how a space economy is feasible with the aggregation of many existing and new technologies into more advanced systems' engineering projects.

## A Resource Challenged Earth

We face unprecedented resource stresses in the 21st Century such as global climate disruptions, freshwater scarcity, expanding energy demands, and the threat of global pandemics. Historically, societies have relieved resource stress by increasing trade, innovating technologically, expanding territorially, regulating, redistributing, making alliances, creating new economic models, training new skills, as well as conducting war. Today we are faced with a choice. Do we continue depleting our already strained resources leading to more regulation, redistribution, alliances, new economics, and war or do we grow our resources using innovation, expansion, new economics, and new skills? Those coming after us in 100 to 300 years will look at their own conditions and judge the early 21st Century generations – how they squandered their resources, remained planet-bound, and created an impoverished planet or how they used the richness of space as their next expansion frontier for the benefit of expanding resources at Earth.

An emerging vision in the early 21$^{st}$ Century is the following – *create the tools for enabling human evolution into space so that life of Earth can be improved*. Capabilities for improving our quality of life can be developed using space assets; paying for that development can be done with a space economy that increases, does not drain, an Earth economy. The Earth-to-space-to-Earth development cycle has already been ad hoc activated with the innovations of New Space; however, the cycle needs a center of gravity for focused actions. We discuss a pathway establishing asteroid mining as the primary activity for frontier expansion that would be aided by Low Earth Orbit (LEO), Moon, and Mars waystations. This pathway advocating a method for humanity to survive climate and environmental crises has been discussed since the expansion of space activity in the 1980's (Hartmann, 1984; Marshall, 1993) and here we offer new innovations for that route.

A technology baseline for developing this pathway is the ability to forecast space weather. Space weather is the transfer of energy from the Sun to the Earth. That energy transfer occurs using photons, particles, and fields. The solar wind, coupled with Earth's magnetosphere, ionosphere, and thermosphere, affects the performance and reliability of space-borne and ground-based technological systems. Changes in space weather conditions cause disruptions in telecommunication, satellite technologies, and power grids as well as pose dangers to those working in space and near-space, ranging from aviation crew to astronauts. All activity off Earth will require a fundamental knowledge of how the energetics of space will affect technological progress. Ensemble, physics-based, data assimilative, machine-learned, empirical, and even statistical modeling, augmented with continuous streams of real-time data, will be the basis for space weather forecasting.

## Asteroid Mining: A Space Gold Rush

Mining
Asteroid mining can help focus a sustainable space economy that would fundamentally revolutionize our approach to resource management and spur technological advancements. Both these activities will help improve life on Earth. Asteroids are solar system bodies replete with metals and minerals. They are a mineral resource that far surpasses our terrestrial reserves.

The idea of mining asteroids for resources is not new. Lewis (1997) and Sonter (1997) both described how asteroids contain a variety of minerals such as iron, nickel, cobalt, and even platinum group metals. AstroForge is an example company making strides into the space mining rush. The company recently announced plans to launch two missions in 2023, i.e., one to refine platinum from a sample of asteroid-like material and another to find an asteroid near Earth to mine.

Furthermore, water found on asteroids could be used not only for life support but also dissociated into hydrogen and oxygen for use as fuel and for a breathable atmosphere (Elvis, 2012). The clear advantage in mining asteroids vs. planets is that less energy is used fighting planetary gravity wells for transferring the ore to processing centers. In addition, the deposits are already accessible near the surface, thus saving energy and time in lieu of deep-mine excavations.

The value of the asteroid belt is analogous to the abundant gold and silver deposits that spurred the mid-1800's American West Gold Rush. For example,

- **Asteroid 16 Psyche** is a large M-type asteroid and has been studied by astronomers in visible and infrared wavelengths since its discovery on 17 March 1852. Psyche might be the partial core of a shattered planetesimal and is a unique metallic asteroid belt object composed of iron and nickel. Estimates suggest that the iron alone in 16 Psyche could be worth around $10,000 quadrillion assuming the volume of material within its 200 km diameter and the current market value of iron (Elkins-Tanton, et al., 2014). NASA's Psyche mission is scheduled for launch in October 2023.
- **Asteroid 162173 Ryugu** is a near-Earth asteroid that was studied by Japan's Hayabusa2 mission. It contains significant amounts of water and organic materials with some suggestions of a value in its materials as high as $82 billion.
- **Asteroid 241 Germania** is a M-type (metallic) asteroid belt object believed to be so rich in precious metals that its value could reach into trillions of dollars. It was discovered on 12 September 1884 and is classified as a B-type asteroid composed of dark, carbonaceous material. The Planetary Resources venture backed by Avatar director James Cameron, which aims to mine asteroids for their wealth, estimates that 241 Germania has $95.8 trillion of mineral wealth inside it, i.e., nearly the same as the annual GDP of the entire world.
- **Total Asteroid Belt** value has been estimated by John S. Lewis in his book "Mining the Sky" (1997). He suggested that the total asteroid belt value might exceed $100 billion for each of Earth's six billion inhabitants (as of the book's publication nearly 3 decades ago). He considered a range of elements and compounds that could be present.

These valuations are purely speculative and are based on several assumptions, including market value of the materials on Earth, the complete extraction and return of the materials, and the absence of any market saturation effects. Realistically, the costs associated with mining, processing, and returning such materials from asteroids are also incredibly high and not fully understood. Thus, these figures are interesting potentialities rather than realistic expectations.

Lithium

Certain types of asteroids, particularly S-type (stony) asteroids, are believed to contain lithium. Lithium, which is used in battery technology, is of particular interest for the power industry. Lithium comes from brine and hard rock. Brine deposits are found in salt lakes and its harvesting there is a common method of extraction, although it yields a lower grade lithium. Hard rock mining requires geological surveys and drilling through rock, which increases costs but results in higher grades of ore. Half of the estimated global lithium resources are in the salt flats of Bolivia, Chile, and Argentina although Australia is the largest producer. China, Portugal, and southern Africa also mine lithium and there are U.S. lithium deposits in Nevada pending possible mining approval.

The question of whether lithium exists in the asteroids is important. Ceres, a dwarf planet, and the largest object in the asteroid belt is likely an ideal place to begin the search for lithium. NASA's Dawn spacecraft visited Ceres in 2015 and discovered bright spots all over the surface. Scientists determined that these are enormous deposits of salt, mixed with a bit of water ice. Recent studies (De Sanctis et al., 2020) concluded that the bright spots such as those in Occator Crater come from an extensive subsurface brine layer that gets partially excavated during impacts. These findings make the Ceres Occator Crater salt flats a prime location for studying the possibility of past or even present extraterrestrial life as well as a possible location for lithium.

We can easily conceive of an eco-system where humans have expanded into the asteroid belt to mine minerals and where sustained growth exists beyond, and independent of, Earth. The economic and technological promise of growth from asteroid mining could be transformative, but it also presents large technical, economic, and ethical challenges.

## Technical Feasibility and Infrastructure Development

Technological foundation begins with space weather

With growing advancements in space exploration and a goal of mining the asteroids, the importance of forecasting space weather increases. For example, spacecraft traveling into deep space or setting up camp on asteroids need to be forewarned of incoming solar storms or radiation events. Depending upon size, these events could add up to an order of magnitude more dose above an existing radiation background. Radiation dose can damage delicate electronics and pose significant health risks to astronauts. While Earth has a magnetosphere and an atmosphere to protect itself from most radiation, asteroids as well as the craft journeying to them lack these protective features.

The magnetic fields and radiation environments of asteroids vary widely. Data from missions like NASA's Galileo and NEAR Shoemaker missions, ESA's Rosetta mission, and Japan's Hayabusa missions have provided some insight. Galileo's encounter with Ida/Dactyl found a binary system unprotected from the harsh solar wind (Figure 1). The NEAR Shoemaker mission to asteroid 433 Eros found that Eros does not have a global magnetic field (Acuña et al., 2002). This discovery supports the idea that most asteroids do not have a significant magnetic field. The Rosetta mission to comet 67P/Churyumov–Gerasimenko revealed an extremely complex and dynamic environment, with plasma interactions driven by the solar wind, outgassing cometary material, and the comet's weakly magnetized nucleus (Goetz et al., 2019). For successful asteroid mining operations, we must be able to understand and predict these conditions.

Accurate space weather forecasts will be a key to protecting the space-based infrastructure and people that are necessary for asteroid mining. This includes not just the mining equipment on the asteroids themselves, but also the transportation systems needed to get to and from the asteroids as well as the communication systems that link these operations back to Earth.

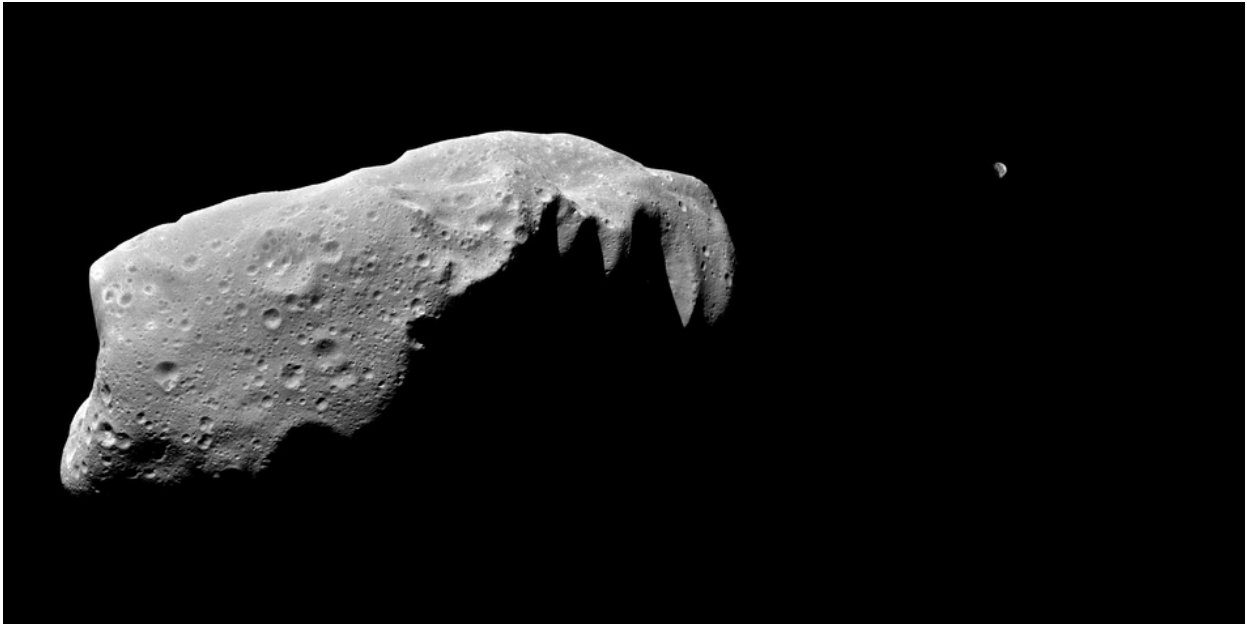

Fig. 1. Ida asteroid and its moon Dactyl as photograph by the Galileo spacecraft (NASA, August 28, 1993).

Scope of infrastructure needs

As we follow the pathway of expansion into space, Low Earth Orbit (LEO), the Moon, and Mars become the key waypoints to the asteroids. They will provide the fundamental logistical hubs with some gravitational benefits if humans are to be involved. They would play an analogous role to the North American Midwest towns along the Mississippi river and the Great Plains two centuries ago. In the 19th Century, waypoints became major population hubs over the course of decades and a similar process would likely occur in LEO and at the Moon and Mars later this century if asteroid mining became the goal of economic activity rather than simply going to the Moon or Mars for their own sake.

Becoming a commercial and industrial hub requires building an infrastructure. That infrastructure needs constant communication and transportation supply lines for expanding habitation, food/water, power, and manufacturing as well as for establishing distribution centers, i.e., an echo of our current terrestrial systems (Crawford, 2015). We have started that infrastructural expansion in the early 21st Century but a discussion of the current Technology Readiness Level (TRL) for these infrastructural technologies is far outside the scope of this paper. Understanding the environmental impact of asteroid mining is an important activity but must also be discussed in other papers.

Looking at our global technological level in 2023, a range of resources are at our disposal. We see the emergence of a new golden age in space exploration and the strengthening of space activity from both government and commercial players. New economic and technological sectors that were previously nonexistent are emerging, such as the building of spaceports, satellite manufacturing, new rockets, space traffic management, space tourism, and commercial space weather forecasting. We see vehicles being built to carry a hundred passengers to Mars and, from a defense perspective, we understand the emergence of a concept called space domain awareness. The latter encompasses the detection, tracking, characterization, and understanding of multiple space situational awareness activities so that decision-making can be made. This includes the use of cyber, ground- and space-based systems that can rapidly detect, warn, characterize, attribute, and predict threats to national, allied, and commercial space systems. Thus, the aggregation of existing technologies into new, expanded systems is becoming as important as pure discovery and this trend will play a significant role for improving life on Earth.

Radiation challenge

Despite these advancements, we are still confronted with enormous technological challenges. A primary one, not yet solved, is that we must be able to safely inhabit asteroids and other celestial bodies, as well as to travel for extended periods and arrive at those deep space locations. To do this we must mitigate an extreme radiation hazard and create effective radiation shielding mechanisms. Active, real-time radiation monitoring systems such as the Automated Radiation Measurements for Aerospace Safety (ARMAS) system from Space Environment Technologies are used in multiple types of craft in the atmosphere through LEO space and to the Moon (Mertens and Tobiska, 2021). However, active radiation protection in all space environments needs further innovation (Durante and Cucinotta, 2008) and there is much work to do in this area.

Closed-loop life support

Another set of unsolved health-related challenges are the effects of long-term living in space as revealed in the NASA twins' study (Garrett-Bakelman et al., 2019). The effects of microgravity on the human body along with mental health concerns in closed environments are important issues. In microgravity, for example, the bones and muscles weaken in a condition like osteoporosis and muscle atrophy. The immune system alters, and fluids shift towards the head, which can affect vision. There is an increased risk of kidney stones from higher levels of calcium in the bloodstream and cardiovascular health can be affected with potential issues such as heart rhythm problems and decreased overall fitness level. Mental health concerns include the effects of isolation and confinement during long-duration space missions, which can cause behavioral issues and psychiatric disorders. Astronauts may experience anxiety, depression, and sleep disorder from a high-stress environment, workload, disrupted circadian rhythm, and isolation from loved ones. The MELiSSA (Micro-Ecological Life Support System Alternative) ESA project showed that a closed-loop life support system must maintain an environment where air, water, and food are continually regenerated while recycling waste products, managing carbon dioxide, and stabilizing temperature levels.

Failures could lead to a shortage of vital resources and pose significant health risks. Closed-loop biological systems may be highly susceptible to bacterial, viral, fungi, insect infestations with results of rapid destruction of hydroponic gardens or the development of mutated viral infections that could spread through a community.

Creating an artificial bio-diverse ecosystem is still far beyond our current capabilities. Mitigation of these risks and issues is a significant area of research for space agencies worldwide, and while these challenges exist, the advancements in technology and our understanding of the human body in space are continually improving the approach we use to solve these issues.

Space Debris

On the entry into space through LEO end of the infrastructure buildout, space debris is becoming a potential barrier to space expansion. By 2023, the European Space Agency (ESA) estimated that there are over 130 million pieces of debris smaller than 1 mm – 1 cm, about 1 million 1–10 cm, and 37,000 of pieces larger than 10 cm. Even tiny particles can cause significant damage due to their high orbital speeds and 1 g of debris colliding with an object releases an equivalent to 40 g of TNT in energy. With an increasing number of satellites being launched through large satellite constellations, e.g., SpaceX's Starlink, the quantity of potential debris will increase substantially with the number of objects in LEO likely to triple over the next 2–3 years. As of 2023, SpaceX alone, for example, has deployed over 4000 Starlink satellites (Figure 2). The concern is that we are approaching the Kessler Syndrome where the density of objects in low Earth orbit increases as collisions create more and more debris. This leads to a cascading effect where further collisions become increasingly likely, and the result potentially makes some LEO orbital regions unusable.

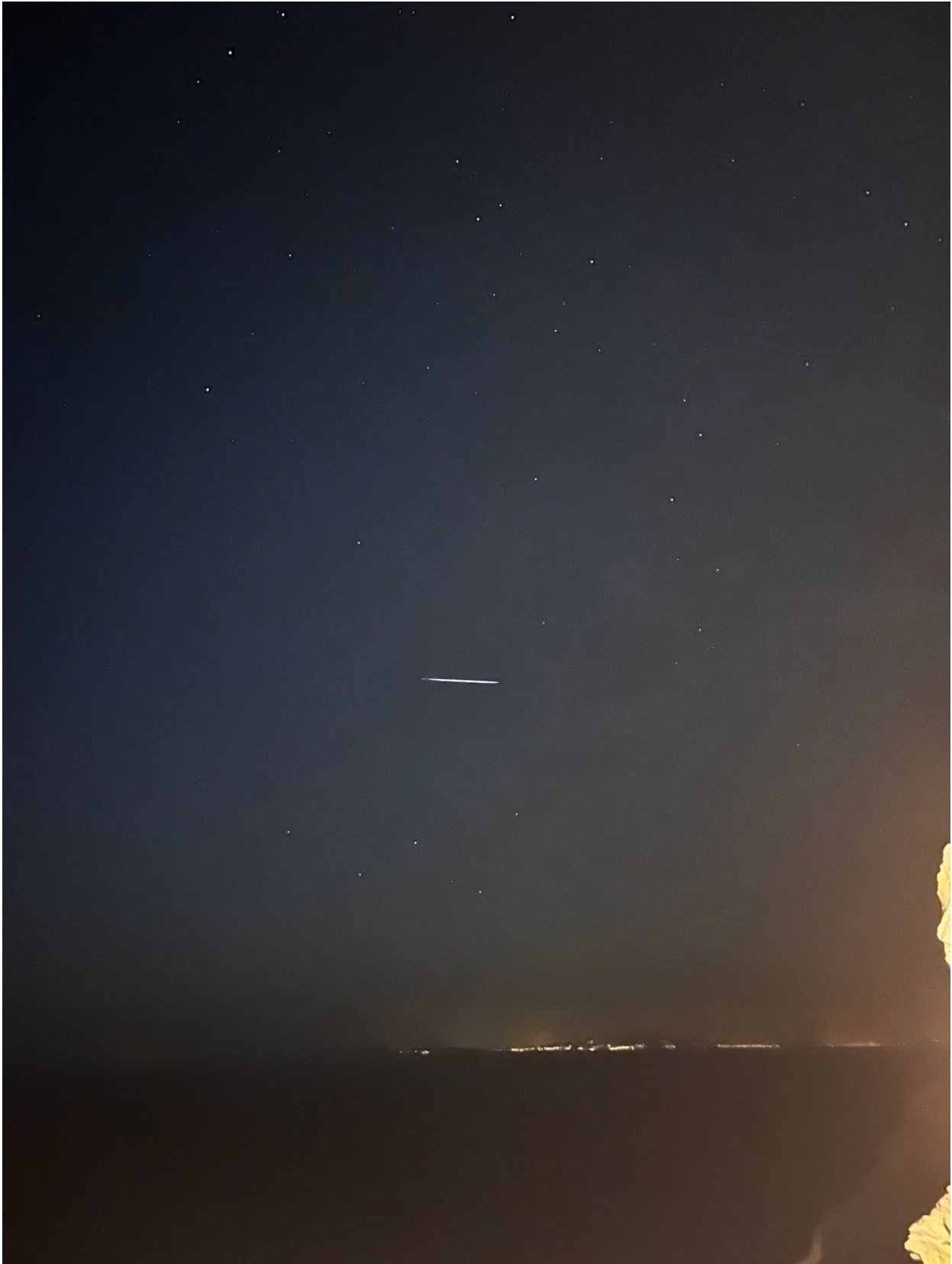

Fig. 2. Post-launch LEO Starlink train (the string of approximately 60 satellites) near 200 km altitude approximately two orbits after launch on June 22, 2023 photographed above the Aegean Sea NW of the island Folegandros (lights of Sifnos can be seen on the horizon) (M. Afzaal, June 22, 2023).

Space debris not only poses risks to expensive infrastructure in space but also to the safety and sustainability of future space activities. It may be possible to "grab and drag" a few hundred large LEO objects greater than 1 meter in size and de-orbit them. For example, the e.Deorbit mission by ESA planned for the mid-2020's is designed to remove a single defunct satellite and is estimated to cost around $300 million. However, the existence of over a million small objects down to 1 cm in size poses a separate, distinct, and unsolved collision hazard. That debris population accounts for most of the objects that could potentially cause catastrophic damage to an active space vehicle.

For the mitigation pathway of small debris objects, innovative concepts such as the "bird seed" removal method may be possible, which suggests deorbiting smaller debris pieces using photons directed from space or ground lasers into the direction opposite the small object's velocity vector to remove orbital energy and decrease the object's lifetime. This and other methods will require rigorous validation and refinement which has not yet started in earnest (Liou, 2011).

Large space structures

In between these two ends of the infrastructure domain lies the task of developing structures comparable to the dimensions of the International Space Station (ISS) and larger. The construction of large, pressurized locations in LEO, GEO, the Moon, Mars, and in the asteroid belt demands substantial enhancements to our current technology and engineering competencies. These were outlined early in the space age by Bekey and Herman (1985). Encouragingly, there are promising commercial ventures now working to create non-government space stations in LEO before 2030.

For deep-space exploration and habitation, it is possible that recent advances in AI and robotics could be leveraged to create less hazardous and cost-effective options related to long-duration exposure in the space environment. Robotic mining may be cost-effective and is certainly one pathway to consider. Robotic mining machines could extract minerals from an asteroid, transfer the ore to a local robotic smelting platform, and then transport the refined ore to a manned near-Earth facility that further packages it for distribution. This example of aggregated technologies is still not mature.

How a space economy might improve life on Earth

An interesting example of how life on Earth might be improved using space assets is the "space water" concept (Tobiska, 2009); it encompasses Earth to Moon technology elements (Figure 3). The "space water" concept was popularized a decade and a half ago in its association with GEO solar power satellites as an energy source for seawater desalination in the arid Southwest U.S. The idea revolves around integrating and aggregating different technologies starting with geosynchronous Earth orbit (GEO) power satellites that can potentially beam generous power to Earth from GEO.

In this concept, beamed power to Earth could be used for multiple purposes: *i)* to desalinate sea water, which is an energy intensive process (reverse osmosis or distillation), for use by agriculture and *ii)* to power a magnetic rail-gun system for ecologically lofting packets of frozen water back into space for use as fuel and breathable oxygen. This concept is a reinvented version of the 20$^{th}$ Century California aqueduct and could constantly supply water to LEO and the Moon as well as drought-plagued locations on Earth. In the space use of water, frozen water packets caught in LEO via orbital rendezvous could then be resent via rail-gun launchers onto the Moon. The feasibility of this example concept is still theoretical since initial packet velocity to reach orbit, the thermodynamics of packet transit through the atmosphere, and even the mechanics and energy needed for rail-gun operations have not yet matured. Potentially, energy could be supplied by beamed power from GEO.

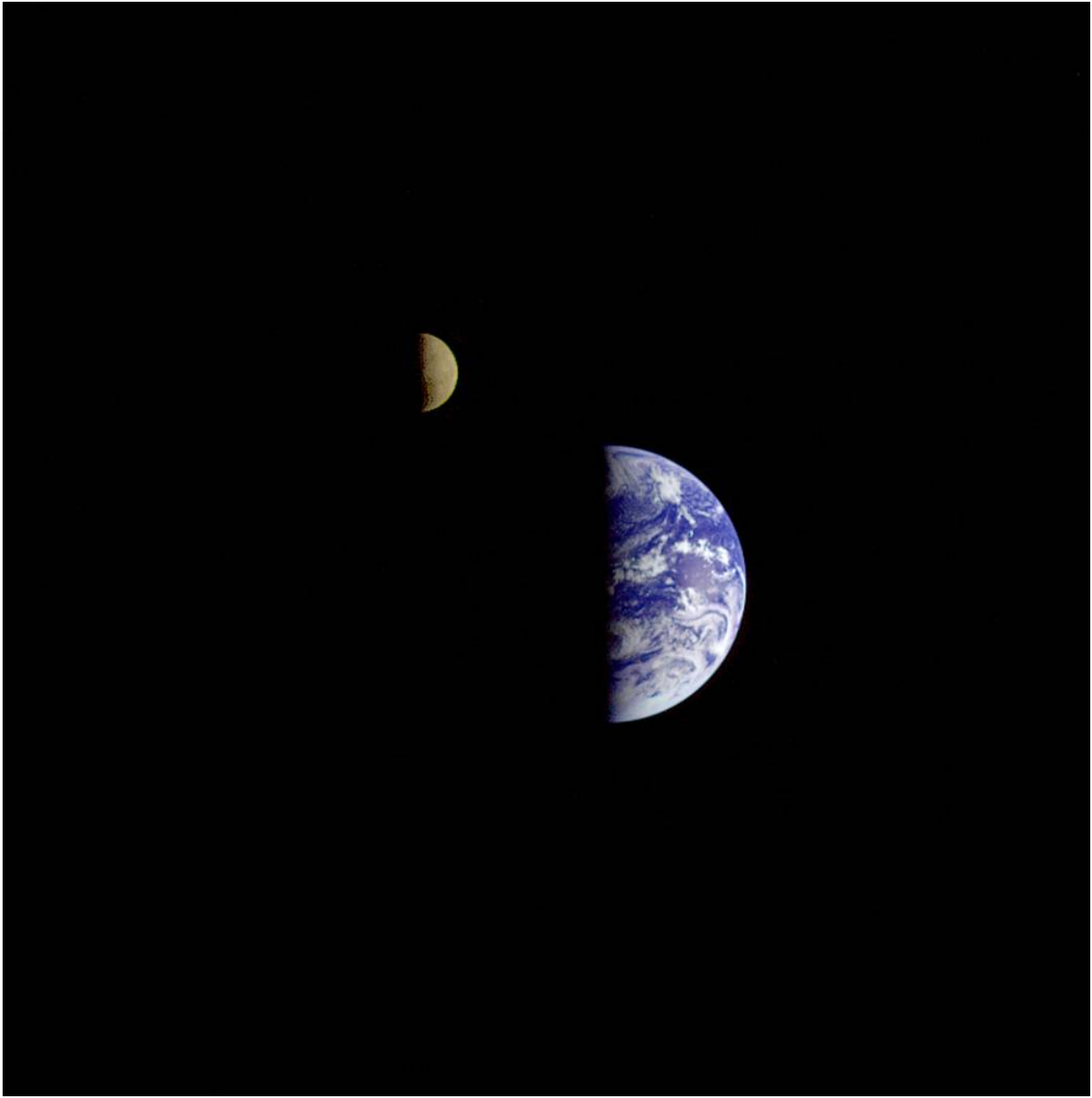

Fig. 3. Earth-Moon system photographed by the Galileo spacecraft Earth-2 flyby (NASA, December 8, 1992).

Other uses of beamed power in space can be envisioned. Transmitted power could be sent to the lunar surface during the Moon's 14-day nights. Beamed power, which could supply LEO and Earth systems, could be time-shared to provide photons for slowly decelerating small debris. In the latter use-case small debris may be more quickly brought to lower altitudes letting the atmosphere deorbit the targeted small debris pieces (diameters of 1–10 cm) sooner via "bird seed" removal. The sidelobes of the power beams, of which an infrared laser may be the best candidate, could potentially be used for directed, light-based communications. Radiation management may imaginatively be addressed with beamed power to space vehicles and structures. In this use-case, miniature magnetospheres might be created around the structures so that the bulk of lower energy charged particles, which are the substantial component of space radiation, can be deflected and removed by an artificial magnetic field. Beamed power has been suggested for propelling solar sails, which has been a concept for decades. Finally, a focused beam could provide ample energy for achieving high temperatures for smelting ore.

## Economic, Geopolitical Viability and a U.S. National Dialogue

The debate surrounding a space economy

There exists considerable debate around the concept of a space economy where the focus is asteroid mining and where LEO, the Moon and Mars are not the endgame but waypoints.

First, there is what might be called techno-pessimism, which views technology as a source of harm or disruption and is often skeptical about the net benefits that technological progress can bring. This perspective (Moore et al. 2022) points to the lack of certainty about potential benefits for Earth. They cast doubt on many of the benefits potentially deriving from space with the arguments that: "(1) other planets are not life-supporting and artificially-supported life might be more vulnerable than on a contaminated planet Earth; (2) the time-frame for geological (or any) exploration of other planetary bodies is greater than the time-frame of extreme environmental change on Earth; (3) the intrinsic value of extra-terrestrial environments is marginalised in favour of anthropocentric concerns; and (4) there is no legal, ethical or societal consensus as to how to share or value other planetary bodies, or to decouple off-Earth extraction from colonial patterns of extractivism. The result is tension between the ecocentric perspective of planetary sustainability and anthropocentric visions of multi-planetary resource acquisition for Earth-based societies."

Despite the inconsistencies of these arguments with the current breadth of space activity, this sentiment partially derives from the view of "why spend Earth's resources to go to space when there are problems on Earth not yet solved." Indeed, this author strongly advocated that view in the 1960's and 1970's. Underpinnings of this view that support techno-pessimism come from: *i)* strong environmentalism with the view that resources used for space exploration could be better used for preserving and restoring Earth's environment; *ii)* social justice/humanitarianism where the importance of addressing social issues such as poverty, hunger, healthcare, and education deserve the bulk of resources spent on space exploration; and *iii)* economic pragmatism where the resources spent on space travel might not yield an adequate return on investment. These perspectives aren't mutually exclusive, and a person might subscribe to several of them.

In the intervening decades, we now know that there are several counterarguments to an overly techno-pessimist stance. Those arguments see long-term benefits of space exploration for scientific knowledge/technological advancement, for resource acquisition, and for survival of humanity. One of the most compelling arguments against techno-pessimism is the undeniable reality of innovation and progress. Technological innovation has brought immeasurable benefits to humanity, including improved health care, increased life expectancy, more efficient communication, and access to vast stores of information. Adaptability resulting from technological changes can bring about social disruption and displacement, but history has shown that societies can and do adapt over time. The fears associated with the Luddite movement during the Industrial Revolution, for example, proved largely unfounded as new industries and jobs eventually emerged to replace those that were lost. A philosophical argument against techno-pessimism is its trap of technological determinism, i.e., the belief that technology shapes society in a particular and inevitable way while overlooking the ability of human beings to shape and direct technological development. Misunderstanding and fear of the unknown can be a basis for techno-pessimism and is fueled by a lack of understanding or fear of the unknown. Education and demystification of technology for the public is important so they can make informed opinions.

In summary, we do need to acknowledge both the potential benefits and risks of technology. A goal, of course, is striving to guide technological development in a way that maximizes the former and minimizes the latter. To this end, we assume that a resilient, successful space economy beyond Earth must pull its own weight and contribute to the collective abundance of human society.

Economic viability

This discussion brings us down to the financial implications of space exploration and asteroid mining. Resilient economic models that encompass initial investment, potential returns, risk mitigation, and long-term sustainability between the surface of the Earth and the asteroid belt still must be developed. A comparison of the benefits of asteroid mining versus other alternatives like Moon and Mars mining within deeper gravity wells, with comparisons of energy expenditure, potential resource yields, or the readiness of the necessary technologies, are important to understand but are outside the scope of this discussion. While government and commercial space flight financial models may provide some guidance, the unique economic factors in the breadth of activity needed for asteroid mining require further thought to ensure its viability (Elvis, 2014).

Geopolitical viability

Space resource extraction has the potential to significantly reshape geopolitics and global power dynamics. For example, if we look at the case of a rare-earth element like platinum, we know that terrestrial supplies are limited, and its mining is environmentally damaging. South Africa, Russia, and Zimbabwe are among the leading producers, and they exert a significant influence on the global platinum market. If extraction of this metal was successfully established through cost-effective asteroid mining operations, either by international government or commercial sector consortiums, this could lead to distinct geopolitical consequences. These could include:

- **A shift in economic power** as the consortium begins to flood the market with space-mined platinum to effectively control global platinum prices and potentially destabilize economies that are reliant on terrestrial platinum mining. Economic power could shift significantly from current platinum producing countries to those who manage space extraction.
- **New alliances and tensions** could spark a new "space race" for resources, leading to new alliances between nations that have space capabilities. On the other hand, tensions could be created between space-faring and non-space-faring nations, potentially escalating into a more fragmented and conflicted international community.
- **An evolution of space law** since the Outer Space Treaty of 1967 currently forms the basis of international space law but is ambiguous about space mining. The large-scale extraction of space resources would create pressure to evolve these laws, which could be a source of international tension. For example, who gets to mine which asteroids? How will profits be distributed? These questions could lead to disputes, and their resolution would shape future balances of power.
- **New security concerns** as space-based assets for resource extraction could be perceived as potential weapons or as possible targets, introducing a new dimension to international security including a new arms race in space.

The shift of resource extraction from Earth to space would likely become a game-changer for geopolitics, influencing economies, international relations, space law, and security. It would create both challenges and opportunities for the global community.

### U.S. National Dialogue

In the U.S., our government agencies are facing severe budget pressures to contract. Yet, there is also an increased effort to address the importance of space weather as related to activities beyond Earth and to ensure natural hazard mitigation for the continuity of governance. Unless a no-brainer argument can be made as to why space weather should be funded at its current levels, spending cuts will likely occur. This paper is a contributor to that no-brainer argument: *make small investments in a sector so that it, as part of an integrated space economy, can create wealth and jobs on Earth from extracted space minerals and energy.*

The U.S. Government formed the Space Weather Advisory Group (SWAG) from the 2020 law entitled *Promoting Research and Observations of Space Weather to Improve the Forecasting of Tomorrow Act* (PROSWIFT, 2020). PROSWIFT defined the roles and responsibilities of the Federal departments and agencies, codified the White House led interagency working group (i.e., the Space Weather Operations, Research, and Mitigation (SWORM) subcommittee), and directed the National Oceanic and Atmospheric Administration (NOAA), working with SWORM, to establish a SWAG. SWAG was chartered to advise SWORM on a variety of space weather issues in 2021, including the development and implementation of an integrated strategy for space weather (i.e., the National Space Weather Strategy and Action Plan). In June 2022, SWORM advised SWAG that they were initiating an update to the National Space Weather Strategy and Action Plan and tasked SWAG to provide input. That response report was provided in April 2023 to SWORM as "Findings and Recommendations to Successfully Implement PROSWIFT and Transform the National Space Weather Enterprise."

In this report, SWAG identified 25 findings with 56 recommendations which, if implemented, would provide the funding, processes, support, and structure to foster transformative change across the national space weather enterprise. Among the 11 highest priority recommendations was one to *provide and fund critical operational space weather services beyond near-Earth.* It was recommended that NOAA, along with the Department of Defense, NASA, and commercial or international partners, support an expanded in-space architecture to meet growing national needs in these regions. This included fulfilling near-Earth operational needs and expanding funding for space weather services in medium-Earth orbit, geostationary Earth orbit, geostationary transfer orbit, cis-Lunar space, and eventually Martian environments. The SWAG/SWORM activity under PROSWIFT has intellectually moved all three of the major pillars of space weather (government, industry, and academia) forward to explore a space economy out to the asteroids as outlined here.

## Ethical and Legal Considerations

Space resource exploitation carries its own ethical and legal challenges, encompassing issues of property rights, territorial disputes, and potential environmental impacts of asteroid mining. International collaboration would be key to establishing comprehensive ethical guidelines and legal frameworks governing these activities (Jakhu, 2006) and this topic is best discussed outside this paper.

## Risk Management and Mitigation

While technical, economic, and ethical considerations are key to asteroid mining, the development of thorough risk management strategies is fundamentally critical. The identification of potential risks in asteroid mining and the creation of comprehensive contingency plans are vital to assure the sustainable growth of a space economy (Ross, 2001). As a society, we have not yet substantially tackled these issues and this topic is also best discussed outside this paper.

## A Space Economy Pathway

We stand at the brink of a new era. The vision of human evolution into space, based upon a flourishing space economy and underpinned by asteroid mining as its objective, promises to be a revolutionary pathway for tackling Earth's challenges during the 21$^{st}$ Century. The advantages of a space economy for benefitting Earth far outweigh concerns of "why spend money on space while we have unsolved problems on Earth?" Earth's problems can no longer be solved by limiting ourselves to Earth – our society requires a space economy that can accelerate our evolution.

A space economy is feasible and is a matter of aggregating many existing technologies into more advanced systems' engineering projects. A result will be the construction of large space structures supporting habitation, transportation, commerce, mining, and manufacturing activities. It relies on expansion to the Moon and Mars by mid-century and sets the stage for the exploratory mining operations on asteroids in the second half of the century. This is the pathway that uses the expanse, energy, and mineral resources of space to enable improvements to life on Earth without the continued destruction of our planet.

Understanding, as well as being able to predict space weather conditions, is an integral first part of building our technological arsenal. Like the meteorologists who provide critical weather forecasts on Earth, space weather forecasters are key to the safety and success of off-world activities.

We see a promise of unlocking the potential of space for assuring a sustainable future on Earth by continuing to address technical, economic, ethical, and risk management complexities. This is not only our pathway to a prosperous future but also a necessary evolution in enabling humans to venture beyond our terrestrial confines to increase our chances for species survival.

## Acknowledgments

This work was assisted by group of futurists who informally met in June 2023 through the auspices of the Anemomilos Boutique Hotel on the Cycladic Island of Folegandros (Greece). A. Galuten, S. Tobiska, J. Bailey, S. Mutschler, D. Bouwer, and K. Leroux contributed insights throughout. Their discussions explored connections between these concepts of a strategic evolution and disruption for Earth's future, loosely referred to as the "Anemomilos Plan." We thank NASA for the Galileo images of the IDA/Dactyl and the Earth-Moon systems. We thank Mr. Muhammad Afzaal for capturing the Starlink train during this retreat with his iPhone camera, a phenomenon visible in early evenings to those present. The Anemomilos hotel staff graciously provided the location and many amenities that led to the success of this retreat for which we are most thankful.

## Availability Statement

There are no data used for, or specifically created by, this manuscript.

## References

Acuña, M. H., B. J. Anderson, P. Wasilewski, G. Kletetshka, L. Zanetti, and N. Omidi (2002). NEAR Magnetic Field Observations at 433 Eros: First Measurements from the Surface of an Asteroid, Icarus **155**, 220–228.

Bekey, I. and Daniel Herman (1985). Space Stations and Space Platforms-Concepts, Design, Infrastructure, and Uses, AIAA, ISBN: 978-0-930403-01-0.

Crawford, I. A. (2015). Lunar resources: A review. *Progress in Physical Geography*, 39(2), 137-167.

De Sanctis, M. C., E. Ammannito, A. Raponi, A. Frigeri, M. Ferrari, F. G. Carrozzo, M. Ciarniello, M. Formisano, B. Rousseau, F. Tosi, F. Zambon, C. A. Raymond, and C. T. Russell (2020). Fresh emplacement of hydrated sodium chloride on Ceres from ascending salty fluids, Nature Astronomy, 4(8):786-793.

Durante M and FA Cucinotta (2008). Heavy ion carcinogenesis and human space exploration. Nat Rev Cancer. 2008 Jun;8(6):465-472. doi: 10.1038/nrc2391. Epub 2008 May 2. PMID: 18451812.

Elkins-Tanton, L. T., et al. (2014). Journey to a Metal World: Concept for a Discovery Mission to Psyche. 45th Lunar and Planetary Science Conference. 1253.pdf.

Elvis, M. (2012). Let's mine asteroids — for science and profit *Nature* 485, 549 https://doi.org/10.1038/485549a

Elvis, M. (2014). How many ore-bearing asteroids? *Planetary and Space Science*, 91, 20-26. https://doi.org/10.1016/j.pss.2013.11.008.

Garrett-Bakelman, F.E. et al. (2019). The NASA twin's study: a multidimensional analysis of a year-long human spaceflight, Science, 12 Apr 2019, 364, 6436, doi: 10.1126/science.aau86

Goetz, C., B. T. Tsurutani, P. Henri, M. Volwerk, E. Behar, N. J. T. Edberg, A. Eriksson, R. Goldstein, P. Mokashi, H. Nilsson, I. Richter, A. Wellbrock, and K. H. Glassmeier (2019). Unusually high magnetic fields in the coma of 67P/Churyumov-Gerasimenko during its high-activity phase, Astron. Astrophys. 630, A38, https://doi.org/10.1051/0004-6361/201833544

Hartmann, W.K. (1984). Space exploration and environmental issues. Environ. Ethics 6, 227–239.

Jakhu, R. S. (2006). Legal Issues Relating to the Global Public Interest in Outer Space, Journal of Space Law, Vol. 32, Pp. 31-110, SSRN: https://ssrn.com/abstract=2801681

Lewis, J. S. (1997). Mining the Sky: Untold Riches From The Asteroids, Comets, And Planets, Helix Books.

Liou, J.-C. (2011). An active debris removal parametric study for LEO environment remediation. *Advances in Space Research*, 47(11), 1865-1876.

Marshall, A. (1993). Ethics and the Extraterrestrial Environment. J. Appl. Philos. 10, 227–236.

Mertens, C.J. and W.K. Tobiska (2021), Space weather radiation effects on high altitude/latitude aircraft, *Space Physics and Aeronomy, Volume 5, Space Weather Effects and Applications*, eds. J. Coster, P.J. Erickson, L.J. Lanzerotti, Y. Zhang, L.J. Paxton, AGU Books, p. 79, ISBN: 978-1-119-50757-4.

Moore, K.R., J. Segura-Salazar, L. Bridges, P. Diallo, K. Doyle, C. Johnson, P. Foster, N. Pollard, N. Whyte, and O. Wright (2022), The out-of-this-world hype cycle: Progression towards sustainable terrestrial resource production, Resources, Conservation & Recycling, 186, 106519.

PROSWIFT Act (Public Law 116-181; 51 USC §§60601- 60608) 2020.

Ross, S. D. (2001). Near-Earth asteroid mining. *Space Industry Report*, 1-14.

Sonter, M. (1997). The technical and economic feasibility of mining the near-Earth asteroids. *Acta Astronautica*, 41(4-10), 637-647.

Space Weather Advisory Group (SWAG) (2023). Findings and Recommendations to Successfully Implement PROSWIFT and Transform the National Space Weather Enterprise, April 17, 2023.

Tobiska, W.K. (2009). Vision for Producing Fresh Water Using Space Power, AIAA Annual Meeting, AIAA-2009-6817.